\documentclass[a4paper,11pt]{article}
\usepackage{jheppub} 

\usepackage{mathtools}
\usepackage{bm}
\usepackage{multicol}

\preprint{CERN-TH-2025-070}

\title{\boldmath Gravitational Wave with Domain Wall Dominance
}

\author[1]{\small Sungwoo Hong,}
\author[2]{\small Sung Mook Lee,}
\author[3]{\small Qiuyue Liang,}

\affiliation[1]{Department of Physics, Korea Advanced Institute of Science and Technology, Daejeon
34141, Korea}
\affiliation[2]{Theoretical Physics Department, CERN, CH-1211 Gen\`eve 23, Switzerland}
\affiliation[3]{Kavli Institute for the Physics and Mathematics of the Universe (WPI), University of Tokyo, Kashiwa 277-8583, Japan}

\emailAdd{sungwooh@kaist.ac.kr}
\emailAdd{sungmook.lee@cern.ch}
\emailAdd{qiuyue.liang@ipmu.jp}

\abstract{
Domain walls (DWs) can be produced when a discrete symmetry is spontaneously broken, and long-lived DWs can dominate the energy density of the universe.
 In this work, we explore the possibility that a ``domain wall dominant (DWD)'' phase existed in the early universe and ended with DW decay.
 During the DWD phase, the universe undergoes a power-law accelerated expansion of the scale factor and exhibits temporal superhorizon evolution of the relevant frequency modes.
 We show that this can lead to distinct features imprinted on the stochastic gravitational wave (GW) background including the independence of its amplitude from the wall tension.
 Our findings provide a comprehensive framework for evaluating GW emission associated with DWD, leading to distinguishable long-lived DW-induced GWs from other cosmological sources, with significant implications for future GW observatories.
}

\begin{document}

\maketitle

\flushbottom

\section{Introduction}

A comprehensive study of the phases of our universe is one of the central goals of contemporary physics and cosmology. 
In cosmology, the concept ``phase'' may be identified with the notion of an ``epoch'' or ``era'' in cosmological evolution.
An interesting question is whether there were other phases beyond the $ \Lambda $CDM model, and if so, what their characteristic features are and how they can be tested.
Especially, while the CMB is currently the earliest direct probe of the early universe, gravitational wave (GW) background, due to their weak interactions, is more transparent and may offer additional insights beyond the last scattering surface. 
Needless to say, such a new phase must be associated with the physics beyond the Standard Model and could provide valuable information of the structure of fundamental physics at earlier times, or higher energy scales.

Topological defects become natural candidates. In fact, their existence and cosmological production are ubiquitous in theories where a global symmetry is spontaneously broken.
In addition, the advent of generalized global symmetry~\cite{Gaiotto:2014kfa} 
and recent applications in particle physics \cite{Brennan:2023mmt} 
shed light on new theoretical perspectives manifested in the physics of topological defects, e.g., global structure of gauge groups~\cite{Cordova:2023her, Choi:2023pdp} or TQFT couplings~\cite{Kapustin:2014gua, Brennan:2023kpw}.
This also motivates us to explore the possible role of these defects in cosmology and universal/distinctive observational signatures associated with them in more detail.

Often considered objects are domain wall (DW), cosmic string, monopole, or hybrid defects, 
and among them, a domain wall dominant (DWD) era is particularly natural to occur, hence deserves careful study.
The energy density of a DW network is known to evolve according to ``scaling solution'', $ \rho_{w} \sim \sigma / t $, where $ \sigma $ is the surface tension of the domain wall.
This implies that in the presence of stable or long-lived DWs, the universe will inevitably enter the DWD phase.
In fact, this phenomenon is known as the domain wall problem and has been regarded as a serious issue in cosmology.
Consequently, most studies on DWs have focused on mechanisms to destabilize them and the resulting predictions.
In contrast, we explore the possibility of a temporary DWD phase.
Furthermore, we investigate the consequences of this transient phase through gravitational wave signals.
Our analysis builds on the parametrized description of GW emission from domain-wall networks~\cite{Gruber:2024pqh} and on the treatment of causality tails~\cite{Caprini:2009fx,Hook:2020phx} extending these ideas to the case of a temporary DWD phase.

The DWD phase is not only natural but also comes with striking features. 
During a DWD epoch, the universe undergoes accelerated expansion with a \textit{reducing} comoving Hubble radius $(aH)^{-1}$, which freezes superhorizon modes that would otherwise never exit the horizon in standard evolution.
Related to this, the characteristic spectrum of the stochastic gravitational wave background (SGWB) signal is also modified compared to the evolution without the DWD phase, and we show that the amplitude is independent of the wall tension, $\sigma$. 
Overall, these changes in peak amplitude and spectrum make this scenario easily distinguishable from GWs produced from other BSM physics or astrophysical sources, hence providing clear signatures of the existence of DWs and the associated new phase in the early universe.

The rest of the paper is organized as follows.
In Sec.~\ref{section:domain_wall_dynamics},  we review DW dynamics in the radiation dominant (RD) phase and discuss the basic features of the DWD phase.
In Sec.~\ref{sec:GW spectrum}, we study the GW spectrum produced by DW networks, taking into account the new features arising from the DWD phase.
Finally, we conclude in Sec.~\ref{section:conclusion_and_discussions}.

\section{Domain Wall Dynamics and Domain Wall Dominance}
\label{section:domain_wall_dynamics}

\begin{figure}[t!]
    \centering
    \includegraphics[width=0.75\linewidth]{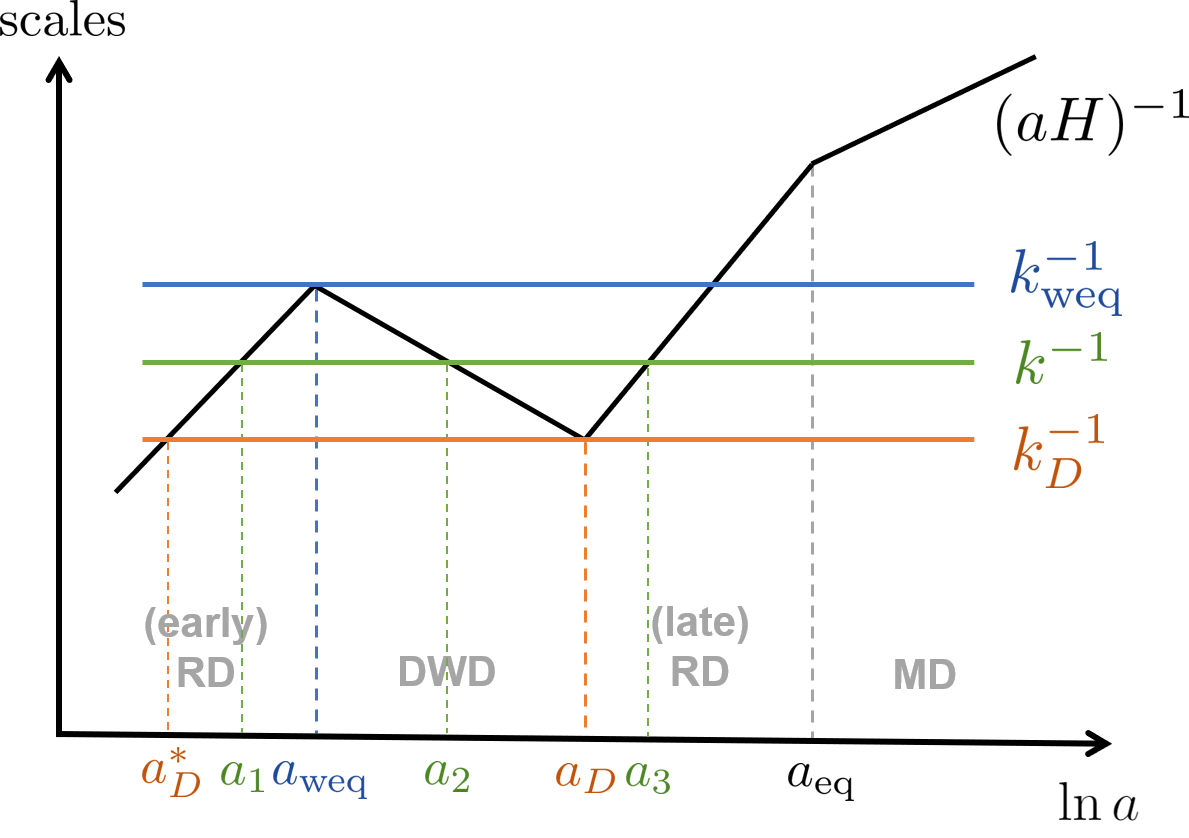}
    \caption{Evolution of comoving scales with the DWD phase.}
    \label{fig:modes}
\end{figure}

We consider the following cosmological history: after inflation, the universe enters the (early) RD era, during which a discrete symmetry breaking phase transition occurs at $t=t_{\rm PT}$, producing DWs. 
The DWD phase begins at $t=t_{\rm weq}$ when the energy density of the walls equals that of radiation, and ends at $t=t_{D}$ when the walls decay into radiation (at $T_{D} \gg {\rm MeV}$).
Afterward, the universe returns to the (late) RD and matter dominated (MD) phase, following standard evolution.
This sequence is illustrated in Fig.~\ref{fig:modes}.

During the RD era, on subhorizon scales, a DW network evolves such that irregularities are stretched out by wall tension and wall interactions.
As this happens, the walls accelerate, and the characteristic length scale $L$ (both curvature radius and inter-wall separation) increases toward the Hubble radius, $H^{-1}$.
It is known that the evolution of the DW network quickly approachs the so-called `scaling regime', where the curvature of the walls relaxes to the size of the Hubble radius, $L\sim H^{-1} \sim t$ and the number of DWs per Hubble volume, $  \sim
\,  L^2 t / H^{-3}$, remains $ \mathcal{O}(1)$.
Therefore, the energy density of DWs can be estimated by $\rho_{w} \sim \sigma / L \sim \sigma/t $. 
Here we assume that the wall tension $\sigma$ is approximately time-independent over the
epoch of interest, neglecting thermal corrections or frictional effects from particle scattering.

This scaling behavior can be inferred by adapting the velocity-one-scale (VOS) model \cite{Kawano:1989mw,Avelino:2005kn,Leite:2011sc} as presented in App.~\ref{supp:VOS Model}, and has also been confirmed from numerical simulations \cite{Hiramatsu:2013qaa,Kitajima:2023cek,Dankovsky:2024zvs}.
Hence, in the presence of long-lived DWs, it is natural to expect that the DW energy density will eventually dominate the universe.
This transition happens at the ``wall-equality'' time, $t_{\rm weq}$, which we estimate by equating the DW energy density with that of the radiation, yielding $ t_{\rm weq} \sim M_{P}^{2} / \sigma$, where the reduced Planck mass is defined as $M_{P} \equiv 1 / \sqrt{8\pi G}$ in terms of the Newton's constant $G$.
We also note that near $ t_{\rm weq}$, the evolution of the DW network is expected to deviate from the scaling behavior.

We now discuss what happens at $t>t_{\rm weq}$, i.e. during the DWD phase.
The evolution of the DW network is expected to deviate from the RD scaling behavior near $t_{\rm weq}$ and the characteristic length scale of a DW network becomes larger than the Hubble radius at $t > t_{\rm weq}$, causing the dominant mode $k \sim L^{-1}$ to undergo superhorizon evolution. 
In fact, superhorizon modes are conformally stretched by the background expansion,
growing as $L \propto a(t)$ as the universe expands. Consequently, $\rho_w \propto 1 / a(t)$ with $ a(t) \propto t^{2} $ which is consistent with Refs.~\cite{Zeldovich:1974uw,Kolb:1990vq,Vilenkin:2000jqa}, and this is also obtained by VOS model discussed in App.~\ref{supp:VOS Model}.
This, in turn, implies that the universe undergoes accelerated expansion and equivalently the comoving Hubble radius decreases, as illustrated in Fig.~\ref{fig:modes}.

While such a DWD phase is often considered a cosmological disaster, and long-lived DWs in the late universe are tightly constrained~\cite{Zeldovich:1974uw,Ferreira:2023jbu}, if the DWD phase is followed by DW decay at $t_{D}$, sufficiently earlier than BBN, it does not necessarily lead to the domain wall problem.
The collapse of DW networks may occur through explicit symmetry breaking effects in the form of a bias potential, the spontaneous nucleation of cosmic strings on the DW world volume~\cite{Kibble:1982dd,Preskill:1992ck,Dunsky:2021tih} or symmetry restoration at low temperatures~\cite{Weinberg:1974hy}.
In this case, the DWD phase can be consistent with current cosmological observations.
Moreover, it may play an important role in diluting unwanted relics present in many BSM scenarios~\cite{Kawasaki:2004rx, Hattori:2015xla}.

In this paper, our main interest is to explore the GW signatures in the presence of an early DWD phase that ends with DW decay at $t_{D}$ (see also~\cite{Bai:2023cqj}). 
As we will discuss later, the GW spectrum is primarily dominated by modes that leave the horizon soon after their emission and is therefore insensitive to subhorizon physics, including the details of the DW decay process.
Consequently, these GW signatures are robust and independent of the specific mechanism by which the DWD phase ends.\footnote{There may be additional contributions from the DW decay process, which we do not take into account here. In general, these contributions populate a higher frequency range.}

\section{Gravitational Wave Spectrum}
\label{sec:GW spectrum}

\subsection{GW Emission Power}

We first discuss the GWs produced by the DW network in the RD era.
We can use the the quadrupole formula ~\cite{Preskill:1991kd,Gleiser:1998na,Hiramatsu:2013qaa,Saikawa:2017hiv} to calculate the  emitted GW power. 
In RD, the quadrupole moment of DWs can be estimated as $ Q \sim M_{w} L^{2} $, where the mass of the wall inside the horizon is given by $ M_{w} \sim  \sigma  L^{2}$.
After reaching the scaling regime, the characteristic length scale follows $L \sim t$.
Using this, the total GW power can be estimated as $
  P_{\rm GW}^{\rm (scaling)} \sim
  G \langle \dddot{ Q }
  ^{2} \rangle \sim  G (M_{w} L^{2} \omega^{3})^{2} \sim  G  \sigma^{2}  t^{2} $, 
and the GW power density is
\begin{align}
   p_{\rm GW}^{\rm (scaling)} = n_{w} P_{\rm GW}^{\rm (scaling)} \equiv \eta_{\rm GW} \frac{G  \sigma^{2}}{t} \ .
   \label{eq:power_scaling}
\end{align}
Here, $ n_{w} \sim L^{-3} \sim  t^{-3} $ denotes the number density of DWs.
According to numerical simulations \cite{Hiramatsu:2013qaa,Kitajima:2023cek,Dankovsky:2024zvs} performed assuming a RD phase, the dimensionless factor $ \eta_{\rm GW} $ is typically $  \mathcal{O}(1) $.

\begin{figure}
    \centering
    \includegraphics[width=0.5\linewidth]{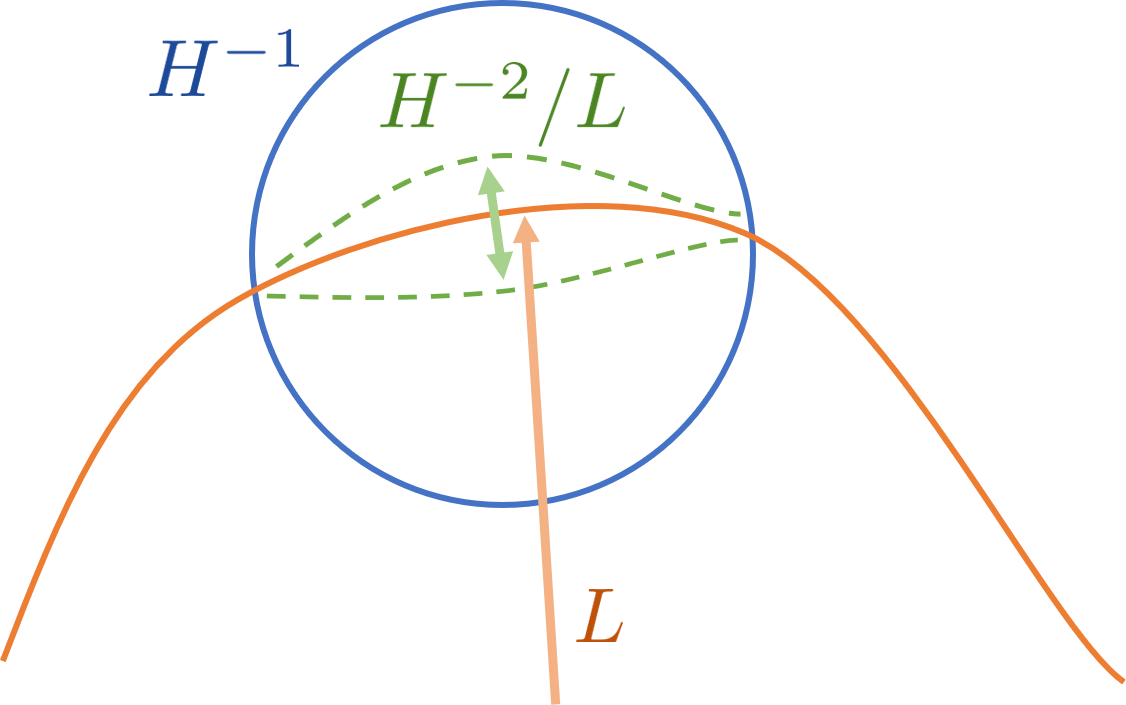}
    \caption{Schematic picture of DW configuration (orange), Hubble horizon (blue) and DW fluctuations within one Hubble patch (green) during DWD phase.    }
    \label{fig:dw_super}
\end{figure}

%
On the other hand, once the universe enters the DWD phase, the situation drastically changes.
While there is only one relevant scale, $ L \sim H^{-1} \sim t $, during the scaling regime in RD, there are three scales we need to consider in DWD: $L, H^{-1}$, and DW fluctuation scale, $H^{-2}/L$. 
Recalling that while $H^{-1} \sim t$, the characteristic length of DW scales as $L(t) \propto a(t) \propto t^2$ in DWD: $L$ increases faster than $H^{-1}$.
Specifically, considering the regime where $L \gg H^{-1} (t)$ at $t > t_{\rm weq}$, the amplitude of the fluctuation is estimated to be $ \sim H^{-2}/L $ (versus $\sim  H^{-1} $ in the scaling regime), as schematically depicted in Fig.~\ref{fig:dw_super}. 
This shows that DW fluctuations are suppressed by a factor $  H^{-1}/L\ll  1$.
Therefore, unlike the scaling regime, the quadrupole moment in a single Hubble patch is expected to be $ Q \sim M_{w} (H^{-2} / L)^{2} \sim \sigma H^{-6}/L^{2} $, where the mass of the wall within the horizon is given to be $ M_{w} \sim  \sigma  H^{-2} $. 
The total GW power emitted within a single Hubble volume is $
  P_{\rm GW}^{\rm (DWD)}
  \sim  G \sigma^{2}  H^{-6} / L^{4} $
assuming the oscillation frequency to be $ \omega \sim H $ since even when
$L \gg H^{-1}$ the GW-emitting dynamics is controlled by causally evolving
structure within each Hubble patch rather than by the global wall scale $L$. Therefore, we expect largely suppressed GW emission during DWD,
\begin{align}
   p_{\rm GW}^{\rm (DWD)} = n_{w} P_{\rm GW}^{\rm (DWD)} \sim G \sigma^{2} H^{-4} / L^{5}   \ll p_{\rm GW}^{\rm (scaling)} \ ,
   \label{eq:power_DWD}
\end{align}
which is strongly suppressed in the DWD regime and will be neglected in the simplified treatment below.
To get the above estimation, we used $ n_{w} \sim (L^2 / H^{-2}) / L^3 \sim H^{2} L^{-1} $, which can be understood by realizing that the number of Hubble patches along the DW surface is $(L^2 / H^{-2})$ and only these patches can contribute to GW signals.
For convenience, we summarize in Tab.~\ref{tab:comparision} the parametric scaling of the quantities entering the GW source estimate in RD and DWD.

\begin{table}[]
    \centering
    \begin{tabular}{|c|c|c|c|c|c|}
    \hline
            &  $a$ & $L$ & fluctuation scale & $p_{\rm GW}$ \\
            \hline
        RD  & $\propto t^{1/2}$  & $\propto t$ & $\sim H^{-1}$ &  $\sim G\sigma^{2}/t$   \\
        \hline
        DWD & $\propto t^{2}$ & $\propto t^{2}$ &  $\sim H^{-2}/L$  & $\sim G \sigma^{2} H^{-4}/L^{5} $ \\
        \hline
    \end{tabular}
    \caption{Parametric comparison of the quantities entering the GW source estimate in RD and DWD.}
    \label{tab:comparision}
\end{table}

Putting everything together, the GW emission power density is expected to be given by $p_{\rm GW}^{\rm (scaling)}$ in Eq.~\eqref{eq:power_scaling} at $t < t_{\rm weq}$, and as $t \to t_{\rm weq}$ it will generally deviate from the scaling solution and eventually will asymptote to $p_{\rm GW}^{\rm (DWD)}$ in Eq.~\eqref{eq:power_DWD} which we approximate as zero.
For our analysis below, we use the following power-law interpolation:
\begin{align}
    p_{\rm GW} = 
    \begin{dcases}
        p_{\rm GW}^{(\rm scaling)} & ( t < r t_{\rm weq} )
        \\
         p_{\rm GW}^{(\rm scaling)} \left(  r t_{\rm weq} / t  \right)^{\alpha} & ( r t_{\rm weq} \leq t < t_{\rm weq} )  
         \\
        0 & ( t \geq t_{\rm weq} ) 
    \end{dcases}\ ,
    \label{eq:power}
\end{align}
for some exponent $\alpha > 0$, and $ r \in [0, 1]  $, which parametrize the time when GW emission starts to deviate from the scaling behavior significantly. Here $t = rt_{\rm weq}$ corresponds to a time when $\rho_w / \rho_r \sim r$.
In principle, $ \alpha $ (or the overall interpolating function) should be determined from simulations.
In the absence of the numerical results on the dynamics of DWs in DWD, we will simply treat it as a free parameter and study its phenomenological effect on gravitational wave power spectrum.

\subsection{Characteristic Scales and Transfer Functions}
\label{sec:Characteristic_Scales_Transfer_Functions}

Given the GW power spectrum at emission, it is useful to introduce the transfer function, $ \mathcal{T}(a_{e},k) $, defined by
\begin{align}
    h_{\lambda} (a_{0}, k) = \mathcal{T} \left( a_{e} , k \right)  h_{\lambda} (a_{e}, k) \, .
\end{align}
where $ a_{0,e} \equiv a(t_{0,e})$ are the scale factors at observation and emission,
respectively, and $k$ is the comoving momentum.
The transfer function encodes the time evolution of the amplitude of the tensor perturbation, or strain $ h_{\lambda} $ ($\lambda = +,\times$) between the time of production ($t = t_{e}$) and observation ($ t= t_{0}$).
As detailed in App.~\ref{supp:tensor_pert}, the strain $ h_{\lambda} $ decays as $ a^{-1} $ on subhorizon and freezes as $a^0$ on superhorizon.
In the absence of the DWD phase, the superhorizon modes at emission ($ k < (a H)_{e}$) begin to decay once they enter the horizon at $ t=t_{*}$ with $ a=a_{*}$, so $ \mathcal{T} = a_{*}/a_{0}$.
On the other hand, for subhorizon modes ($ k > (a H)_{e}$), $\mathcal{T}(a_{e},k) = a_{e}/a_{0}$.\footnote{Since we only need the magnitude of the transfer function, we neglect irrelevant phases in this paper.}

A key feature of the DWD phase is the decreasing comoving Hubble radius.
As a result, modes that would otherwise remain subhorizon can instead spend some time on superhorizon, and the transfer function is modified accordingly.
Moreover, the transfer function can differ depending on the frequency and emission time.

As shown in Fig.~\ref{fig:modes}, there are two relevant comoving momenta given by $ k_{\rm weq} \equiv (a H)_{\rm weq}$ and $k_{D} \equiv (a H)_{D}$ with $ H_{\text{weq}(D)} = \xi_{\text{weq}(D)} / t_{\text{weq}(D)} $, respectively. 
In this work, for numerical evaluations, we use $\xi_{\text{weq}(D)} \simeq 1/2$.
Additionally, modes leaving the horizon at time $ t = r t_{\rm weq}$ (where GW power starts to be suppressed compared to the scaling regime) have
$ k_{r} \equiv k_{\rm weq} / ( 2\xi_{\rm weq} \sqrt{r} )$.

Assuming instantaneous DW decay at $t_D$ into radiation with temperature $T_D$,
the scale $k_D\equiv (aH)_D$ follows from $k_D=a_D H_D$, where
$H_{D}=\sqrt{\pi^2 g_{*}(T_{D})/90}\,T_{D}^2/M_{P}$, and
$a_{D}/a_{0}=(T_{0}/T_{D}) (g_{*s}(T_{0})/g_{*s}(T_{D}) )^{1/3}$ is fixed by
entropy conservation.
Here $g_*(T)$ and $g_{*s}(T)$ denote the effective
numbers of relativistic degrees of freedom for the energy and entropy
densities, respectively.
Similarly, $k_{\rm weq}\equiv (aH)_{\rm weq}$ follows  
from $\rho_w(t_{\rm weq})=\rho_r(t_{\rm weq})$, together with
$\rho_w \propto a^{-1}$ during the DWD phase.
This gives
\begin{equation}
    \begin{aligned}
    k_{D} 
    & \simeq 1.7 \times 10^{-7} \left( \frac{g_{*D}}{106.75} \right)^{1/6} \left( 
 \frac{T_{D}}{\rm GeV}\right)  ~ \text{Hz} \, , \\
    k_{\rm weq} 
        & \simeq 2.1 \times 10^{12} \left( \frac{g_{*}(T_{D})}{106.75} \right)^{2/3} \left( \frac{T_{D}^{3}}{\sigma} 
    \right)  ~\text{Hz} \,. \label{eq:kWD}
    \end{aligned}
\end{equation}
The detailed derivation is given in App.~\ref{supp:Derivation_of_Characteristic_Scales}.

\begin{figure}
    \centering
    \includegraphics[width=0.85\linewidth]{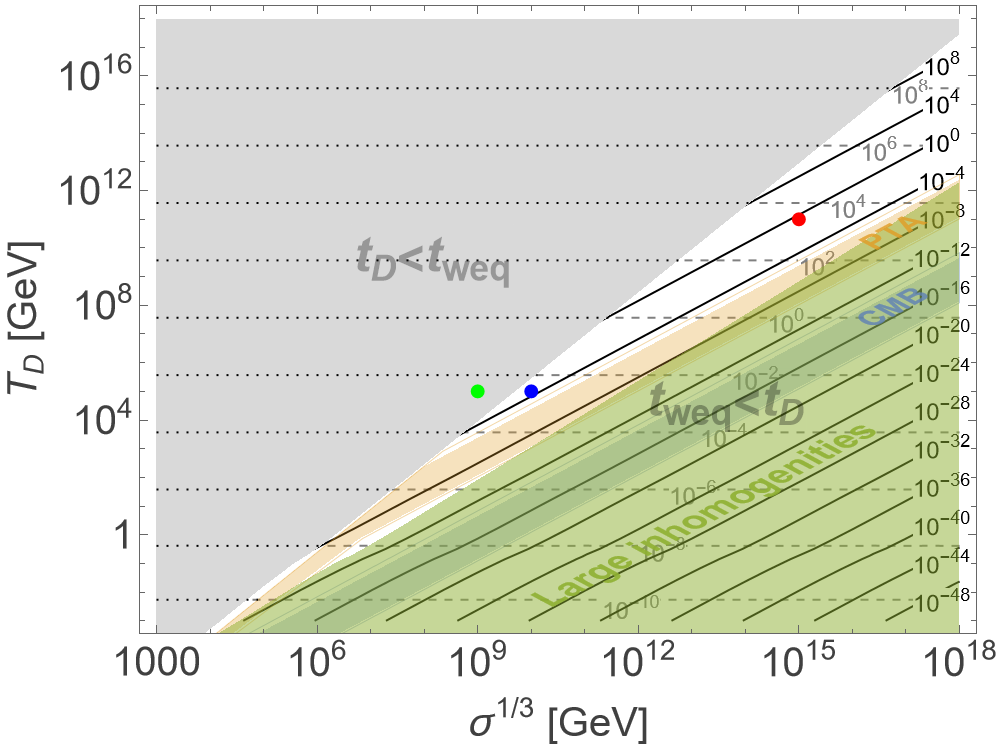}
    \caption{
    The frequencies $ f_{\rm weq} = k_{\rm weq} / (2\pi) $ (black solid) and $ f_{D} = k_{D} / (2\pi) $ (gray dashed) corresponding to the comoving horizon scale at $t_{\rm weq}$ and $t_D$, respectively. 
    The gray-colored region corresponds to $ t_{D} < t_{\rm weq}$, in which DWD phase is not realized.
    The two colored (blue and red) dots denote two example benchmark points: (\textcolor{blue}{BM1}) $ T_{D} = 10^{5} \, \text{GeV} $ and $ \sigma = (10^{10} \, \text{GeV})^{3} $ and (\textcolor{red}{BM2}) $ T_{D} = 10^{11} \, \text{GeV} $ and $ \sigma = (10^{15} \, \text{GeV})^{3} $. 
    The green dot (\textcolor{green}{BM3}) $ T_{D} = 10^{5} \, \text{GeV} $ and $ \sigma = (10^{9} \, \text{GeV})^{3} $ corresponds to a parameter set where the DWD phase does not occur.
    In the light gray region, the dotted lines denote the peak frequency determined by $ T_{D} $.
    The green-colored region is ruled out due to the large inhomogeneities induced on the CMB/LSS.
    The blue- and orange-shaded regions are constrained by CMB \cite{BICEP:2021xfz} and PTA \cite{NANOGrav:2023gor,EPTA:2023fyk}, respectively.
    }
    \label{fig:inhom}
\end{figure}

In Fig.~\ref{fig:inhom}, we show 
contours of  $ f_{\rm weq} \equiv k_{\rm weq} / (2\pi ) $ and $f_{D} \equiv  k_{D} / (2\pi ) $ in $(\sigma^{1/3} \,, T_{D})$ plane, together with several constraints explained later and in App.~\ref{supp:Fluctuation from DW Networks}.
The gray shaded region denotes the absence of DWD phase.\footnote{Not all of these parameters are allowed. For instance, see Refs.~\cite{Saikawa:2017hiv,Ramberg:2022irf,Ferreira:2023jbu} for the observational constraints on DWs without a dominance phase.}
In this case, the peak frequency corresponds to $ k_{D}/(2\pi) $, shown as dotted lines. For illustration, we have chosen three benchmarks, as described in the caption of the figure as well as Tab.~\ref{tab:benchmarks}.

\begin{table}[t]
    \centering
    \begin{tabular}{|c|c|c|c|c|c|c|}
    \hline
         & $T_{D}$ (GeV) & $\sigma^{1/3}$ (GeV) &  $r$ & $k_{D}$ (Hz) & $k_{\rm weq}$ (Hz) & $k_{r}$ (Hz) \\
         \hline
        BM1 & $10^{5}$  & $10^{10}$ & 0.1 & 0.017 & 0.0021 & 0.0066   \\
        \hline
        BM2 & $10^{11}$  & $10^{15}$ & 0.1 & $1.7 \times 10^{4}$ &  2.1  & 6.6 \\
        \hline
        BM3 & $10^{5}$  & $10^{9}$  & - & 0.017 & -& - \\
    \hline
    \end{tabular}
    \caption{Benchmark values and corresponding typical characteristic frequencies. Note that in terms of angular frequency $ f $, $ k = 2 \pi f $.
    }
    \label{tab:benchmarks}
\end{table}

If we consider a specific mode $k  \in (k_{\rm weq},  k_{D}) $ , it crosses the comoving horizon—either exiting or entering— during (early) RD, DWD, or (late) RD.
We define scale factors associated with these crossing points as $a_{1,2,3}$, respectively.
Using the relation between $k$ and $a$, we express these as $ a_1 \equiv a_{\rm weq} \left( k / k_{\rm weq} \right)^{-1} $, $   a_2 \equiv a_{D} \left( k / k_{D} \right)^{2}  $, and $ a_3 \equiv a_{\rm eq} \left( k / k_{\rm eq} \right)^{-1} $.

The transfer function is determined as follows.
For a given emission time $t_e$, we consider cases $k > (aH)_e$ or $k < (aH)_e$.
Further subdivision arises when comparing $k$ to $k_{\rm weq}$ and $k_D$. Whenever $k>(aH)$ (subhorizon), $\mathcal{T} \propto a$, while for $k < (aH)$ (superhorizon), $\mathcal{T}$ remains constant.
First, consider the case $ t_{e} < t_{D}^{*} $, where $t_{D}^{*}$ is defined by $ k_{D} = a H \vert_{t= t_{D}^{*} } $, and $t_{D}^{*} < t_{\rm weq}$ (see Fig.~\ref{fig:modes}).
In this case, the transfer function for modes $ k \geq (aH)_e $ is given by $ \mathcal{T}(a_{e},k) =   a_{e} / a_{0}  $.
Second, for $ t_{D}^{*} < t_{e} < t_{\rm weq}$, we have
\begin{align}
    \mathcal{T}(a_{e},k) = 
    \begin{dcases}
         \left( \frac{ a_{e} }{ a_2 } \right)  
       \left( \frac{ a_3 }{ a_{0} } \right)  &   \left( (aH)_e \leq k <   k_{D} \right) \\
          \left( \frac{ a_{e} }{ a_{0} }  \right)
          &   \left(  k_{D}  \leq k  \right)
    \end{dcases}\ . \label{eq:transfer_2}
\end{align}
Transfer functions for more general cases are provided in App.~\ref{supp:Transfer Functions}.

\subsection{Causality Tail with Domain Wall Dominance}
\label{sec:causality_tail}

To determine the full spectral shape of the SGWB, we need to consider the causality tail in more detail, since it is relevant for the low-frequency part of the spectrum.
For a causal source, the GW spectrum necessarily contains an infrared tail, whose scaling depends not only on the source itself but also on
the subsequent background evolution.
This so-called causality tail describes the low $k$ behavior of the infrared components of a causal source with finite correlation length smaller than the horizon scale, rather than the GWs that are generated acausally outside the horizon.
In an RD universe, it is well known that the causality tail has the spectral shape $k^3$ at low frequency \cite{Caprini:2009fx}, while it can be modified for different background evolutions \cite{Hook:2020phx}.
In the presence of temporary DWD phase, the decreasing comoving Hubble radius allows part of the propagation to occur on superhorizon scales, which modifies the usual infrared behavior as we will discuss.

To make the causality argument explicit, let us consider a localized source $J(\tau)=J_* \delta(\tau-\tau_*)$ at some (conformal) time $ \tau_{*} $ on the right-hand side of Eq.~\eqref{eq:tensor_pert}, and then
follow the subsequent propagation of the corresponding tensor mode:
\begin{align}
    h^{\prime\prime} + 2 \mathcal{H} h^{\prime} + k^{2} h = J_{*} \delta(\tau - \tau_{*}) \, .
\end{align}
Imposing $ h(\tau) \equiv 0  $ for $ \tau < \tau_{*} $ dictated by causality and continuity of the solution, and integrating both sides over an infinitesimal interval ($ \tau_{*} - \epsilon , \tau_{*} + \epsilon$) results in a condition: $ h_{*}^{\prime}(\tau_{*} + \epsilon) = J_{*} $. Hence, the problem of interest is equivalent to solving
\begin{align}
    h^{\prime\prime} + 2 \mathcal{H} h^{\prime} + k^{2} h = 0 \label{eq:h}
\end{align}
with initial conditions
\begin{align}
    h(\tau_{*}) =0 \,, \qquad
    h^{\prime}(\tau_{*}) = J_{*} \,. \label{eq:ic}
\end{align}

The energy density of GW is given by
\begin{align}
\label{eq:rho_GW_x}
    \rho_{\rm GW} (\mathbf{x},\tau) \sim \frac{1}{32 \pi G a^{2}} \langle h^{\prime} (\mathbf{x},\tau) h^{\prime} (\mathbf{x},\tau) \rangle 
\end{align}
and we can write the GW power spectrum as
\begin{align}
\label{eq:Omega_GW_k}
   \Omega_{\rm GW}(k,\tau) \equiv \frac{1}{\rho_{c}} \frac{d \rho_{\rm GW}}{d \ln k} \propto  \frac{k^{5}}{a^2} P_{h}(k,\tau) \,.
\end{align}
where dimensionful power spectrum is defined as $  \langle h (k, \tau) h (k^\prime, \tau) \rangle = 
(2\pi)^{3} \delta^3 (k - k^{\prime})  P_{h}(k, \tau)  $ in momentum space and we used $h^\prime \sim k h$.  An additional $k^3$ comes from the phase space factor appearing when Eq.~\eqref{eq:rho_GW_x} is Fourier transformed into momentum space to get Eq.~\eqref{eq:Omega_GW_k}.
Therefore, the problem of determining the spectral shape turns into figuring out the $ k $-dependence of the power spectrum $P_{h}$, which in turn is the same as finding $k$ scaling of $ h $ by solving Eqs.~\eqref{eq:h} and \eqref{eq:ic}.

Let us consider a superhorizon mode at the time of its generation ($ k \ll \mathcal{H}_{*} $). This is equivalent to the problem of over-damped harmonic oscillator. The velocity quickly (roughly within a single $e$-fold) vanishes and the initial conditions evolve into
\begin{align}
    h^{\prime}(\tau_{*}) \rightarrow 0 \,, \qquad h(\tau_{*}) \sim \frac{J_{*}}{\mathcal{H}_{*}} \,.
\end{align}
The over-damping picture shows that superhorizon evolution has an effect of suppressing the fluctuation. However, there exists a competing effect that enhances the power: namely, on superhorizon, the amplitude is frozen, whereas it decays as $\propto a^{-1}$ on subhorizon. Therefore, an initially superhorizon mode is first frozen until it enters the horizon at $ \tau_{k} $ such that $ \mathcal{H}(\tau_{k}) = k $, and starts decaying as $\propto a^{-1}$:
\begin{align}
    h \simeq \frac{a(\tau_{k})}{a(\tau)} \frac{J_{*}}{\mathcal{H}_{*}} \sin k \tau \,.
\end{align}

For the mode entering the horizon during RD, we can use $ a \propto \tau $ and $ \mathcal{H} = 1/\tau \propto a^{-1}$ to obtain
    \begin{align}
        h \simeq \frac{a(\tau_{k})}{a(\tau)} \frac{J_{*}}{\mathcal{H}_{*}} = \frac{a(\tau_{*})}{a(\tau)} \frac{J_{*}}{k} \propto \frac{1}{k}
    \end{align}
and therefore $P_{h} \propto k^{-2}$ and $\Omega_{\rm GW} \propto k^{3}$, as expected. However, note that this depends on when the mode enters the horizon: RD, MD, or else, e.g.~DWD.
For instance, in the case of MD, $ k = \mathcal{H}(\tau_{k}) = \left( \frac{a(\tau_{*})}{a(\tau_{k})} \right)^{1/2} \mathcal{H}_{*} $ so that $ h \propto k^{-2} $ and $ \Omega_{\rm GW} \propto k $, different from $ k ^{3} $. In Ref.~\cite{Hook:2020phx}, it is shown that for the equation of state $w$, $ \Omega_{\rm GW} \propto k^{\frac{15w + 1 }{3 w + 1}}$.

In the case of domain wall dominance, the situation is a bit more involved, because of the presence of partial subhorizon/superhorizon evolutions during its propagation. For domain wall, source is not localized in time, but rather long-lived. However, recalling that the dominant contribution comes from production at the latest time, we can use the same argument for   that time slice as before.

If $ k < k_{\rm weq}$, the mode enters subhorizon at RD and the above calculation holds, leading to 
\begin{align}
    \Omega_{\rm GW} \propto k^{3} \qquad (k<k_{\rm weq})
\end{align}
On the other hand, for $ k > k_{\rm weq} $, the situation is different. Especially, for modes $ k_{\rm weq} < k < k_{D} $, there are three times when $k$ becomes equal to comoving Hubble radius: $a_{1,2,3}$ as defined in Sec.~\ref{sec:Characteristic_Scales_Transfer_Functions} (see Fig.~\ref{fig:modes}).
Then, the mode enters the horizon during the (early) RD at $a_1$, and suffers subhorizon evolution for a while, then it exits the horizon at $a_2$ and re-enters once again at $a_3$. 

Therefore, the solution becomes
\begin{align}
    h \simeq \left( \frac{a_3}{a(\tau)} \right)^{1} \left( \frac{a_2}{a_3} \right)^{0}  \left( \frac{a_1}{a_2} \right)^{1}  \frac{J_{*}}{\mathcal{H}_{*}} .
\end{align}
As before, $ a_1 =  a(\tau_{*}) \mathcal{H}_{*} / k $ and now we have an additional factor to take into account.
\begin{align}
    \frac{a_3}{a_2}  = \frac{a(\tau_{D})}{a_2} \cdot \frac{a_3}{a(\tau_{D})} = \left( \frac{\mathcal{H}_{D}}{\mathcal{H}_2} \right)^{2}   \left( \frac{\mathcal{H}_{D}}{\mathcal{H}_3} \right)  = \left( \frac{k_{D}}{k}  \right)^{3} \,.
\end{align}
where $\mathcal{H}_i \equiv \mathcal{H} (a_i)$ and we used that $k = \mathcal{H}_i$ for $i=1,2,3$.
Hence,
\begin{align}
     h \simeq \left( \frac{a_3}{a_2} \right) \frac{a(\tau_{*})}{a(\tau)}   \frac{J_{*}}{k} \propto k^{-4}
\end{align}
giving
\begin{align}
   \Omega_{\rm GW} \propto k^{5} \cdot k^{-8} = k^{-3}  \qquad (k_{\rm weq} < k < k_{*}) \,.
\end{align}

We can also generalize the situation for arbitrary equation of states during accelerated phase, $  w < -1/3 $ with scale factor $ a(t) \propto t^{\frac{2}{3(1+w)}}$. Domain wall dominance with $a(t)\propto t^{2} $ corresponds to the case of $ w = - 2/3$. This just modifies the power of $\left( \mathcal{H}_{D} / \mathcal{H}_2 \right) $ term from 2 to $-2/(3 w + 1)$, resulting in $ h \propto k^{- \frac{6w}{1+3w}}$, or $ \Omega_{\rm GW} \propto k^{\frac{5+3 w}{1+3w}} $. For $ w = -2/3 $, this reduces to $ \Omega \propto k^{-3} $ as expected.

\subsection{GW Power Spectrum}
\label{subsec:GW spectrum}

For a given $ p_{\rm GW} $ at a specific emission time $t_{e}$ (see Eqs.~\eqref{eq:power_scaling} and \eqref{eq:power}), the differential power of GWs is assumed to be separable:
\begin{align}
    \frac{ dp_{\rm GW} }{d \tilde{k}_{e}} (t_{e}, \tilde{k}_{e} )= p_{\rm GW}(t_{e}) \mathcal{P}( \tilde{k}_{e} ) \ .
\end{align}
Here $\mathcal{P}( \tilde{k}_{e} )$ represents the probability density function (PDF) that satisfies $\int_{0}^{\infty}d\tilde{k}~ \mathcal{P}(\tilde{k}) = 1 $.
$ \tilde{k}_e $ denotes the \textit{physical} momentum at emission related to the comoving momentum through $ \tilde{k}_e = k_e / a $.
We note that $p_{\rm GW}(t)$ denotes the GW power density integrated over emitted physical momentum, while $d p_{\rm GW}/d\tilde{k}_e$ is the corresponding differential emission spectrum.

We consider a simple power-law PDF of the form $\propto \tilde{k}^{-\nu}$ ($\nu > 1$) as discussed in~\cite{Gruber:2024pqh}, 
\begin{align}
    \mathcal{P}(\tilde{k}_{e}) = \frac{\nu - 1}{\tilde{k}_{\min}^{- \nu + 1}(t_{e})} \tilde{k}_{e}^{-\nu} \Theta(\tilde{k}_{e}-\tilde{k}_{\min}(t_{e}))\ ,
\end{align}
where $ \tilde{k}_{\min}(t) \simeq  H(t)  $.
This does not mean that there are no GWs at lower frequencies below $\tilde{k}_{\rm min}$.
In fact, there will be an irreducible white noise in the form of causality tail at lower frequency, as we discuss later.
To obtain the energy spectrum observed today, GWs produced from the time of DW formation to its decay have to be integrated.
In addition, each GW mode has to be evolved with a proper transfer function, and when this is done, evolution of background spacetime should also be taken into account.
For instance, for subhorizon modes, the Boltzmann equation for GW energy density is given by $ d \rho_{\rm GW}/ dt + 4 H \rho_{\rm GW} = p_{\rm GW} $ and the information of the background evolution is included in the scale factor, which should be integrated along with the emission power to take into account the effects of the dilution. 
Explicitly, the observed energy density spectrum is
\begin{align}
   \left. \frac{d\rho_{\rm GW}}{dk} \right\vert_{0} =
   \int_{t_{\rm PT}}^{t_{0}} dt~a(t) \left\vert \mathcal{T} \left( a(t), k \right) \right\vert^{2}   p_{\rm GW}(t)  \mathcal{P} \left(  \frac{k}{a } \right) 
   \label{eq:master} 
\end{align}
where we replace $ t_{e} $ to $ t $ (hence $a_{e}$ to $ a$) and $  k $ is the comoving momentum corresponding to the observed frequency today.
Hereafter, we set $ a_{0} \equiv 1 $. 
Since GWs generated earlier are more suppressed due to redshift, we can safely set $t_{\rm PT} = 0$.
The quantity frequently used to characterize cosmological GW backgrounds is $  h^{2} \Omega_{\rm GW} (k) = \frac{h^{2}}{\rho_{c,0}} \left. \frac{d\rho_{\rm GW}}{d \log k} \right\vert_{0} 
$, where $ \rho_{c,0} = 3H_{0}^{2} / (8 \pi G) $ is the critical energy density of the universe today.

\subsubsection{GW from DW only with RD}

Assuming a power-law PDF, GW produced during RD without a DWD phase (hence taking $ \mathcal{T} \propto a $) takes the form $ h^{2} \Omega_{\rm GW} (k)  \propto k^{-\nu+1} $ for $\nu < 5$.
Numerical simulations with $ \nu = 2 $ suggest $ \eta_{\rm GW} \simeq 0.7 \pm 0.4 $ \cite{Hiramatsu:2013qaa}.

It is worth commenting on further results from the numerical simulations.
Early studies claimed $ \Omega_{\rm GW} \propto k^{-1} $ at high $ k $ \cite{Hiramatsu:2013qaa} corresponding to $ \nu = 2 $.
More recent works, however, have shown that $ \Omega_{\rm GW} \propto k^{-1.8 \sim -1.7}$ \cite{Dankovsky:2024zvs}, which requires larger $ \nu $ in our parameterization, and exhibits bumpy/plateau features at higher $k$, likely dependent on initial conditions \cite{Kitajima:2023cek,Dankovsky:2024zvs}.

\subsubsection{GW from DW with DWD}

In the presence of a DWD phase, the spectrum of GW is significantly modified.
Performing the integration given in Eq.~\eqref{eq:master} with $t_{e} \in (0, t_{\rm weq})$, and neglecting GW production after $ t_{\rm weq} $, we find for $ \nu = 2 $ and $ k_{r} < k_{D}$: 
\begin{align}
    h^{2} \Omega_{\rm GW} (k)  
    \simeq \begin{dcases}
         \tfrac{2r^{\alpha}}{3 - 2 \alpha} \mathcal{A} \left[ \left( \tfrac{k}{ k_{\rm weq}} \right)^{ \scriptscriptstyle -7}
       - \left( \tfrac{k }{ k_{\rm weq}} \right)^{\scriptscriptstyle 
 -10+2\alpha} 
       \right] & ( k_{\rm weq} < k < k_{r} ) \\
       \tfrac{6r^{\frac{7}{2} + \alpha} - 4 r^{5} \alpha }{9 - 6\alpha} \mathcal{A}  
       \left( 
       \tfrac{k}{k_{r}} \right)^{\scriptscriptstyle  -7} & ( k_{r} < k < k_{D} ) \\
       \tfrac{6r^{\alpha} - 4 r^{\frac{3}{2}} \alpha }{9 - 6\alpha}  \mathcal{A} 
       \left(  \tfrac{k_{D}}{k_{\rm weq}}\right)^{\scriptscriptstyle  -7}
       \left( 
       \tfrac{k}{k_{D}} \right)^{-1} & (k > k_{D}) 
    \end{dcases} 
    \label{eq:spectrum_2} 
\end{align}  
where $ \mathcal{A} = 8\pi h^{2} \eta_{\rm GW} a_{\rm eq}^{4} H_{\rm eq}^{2} /(3H_{0}^{2}) $.

\begin{figure}[t!]
    \centering
    \includegraphics[width=0.95\linewidth]{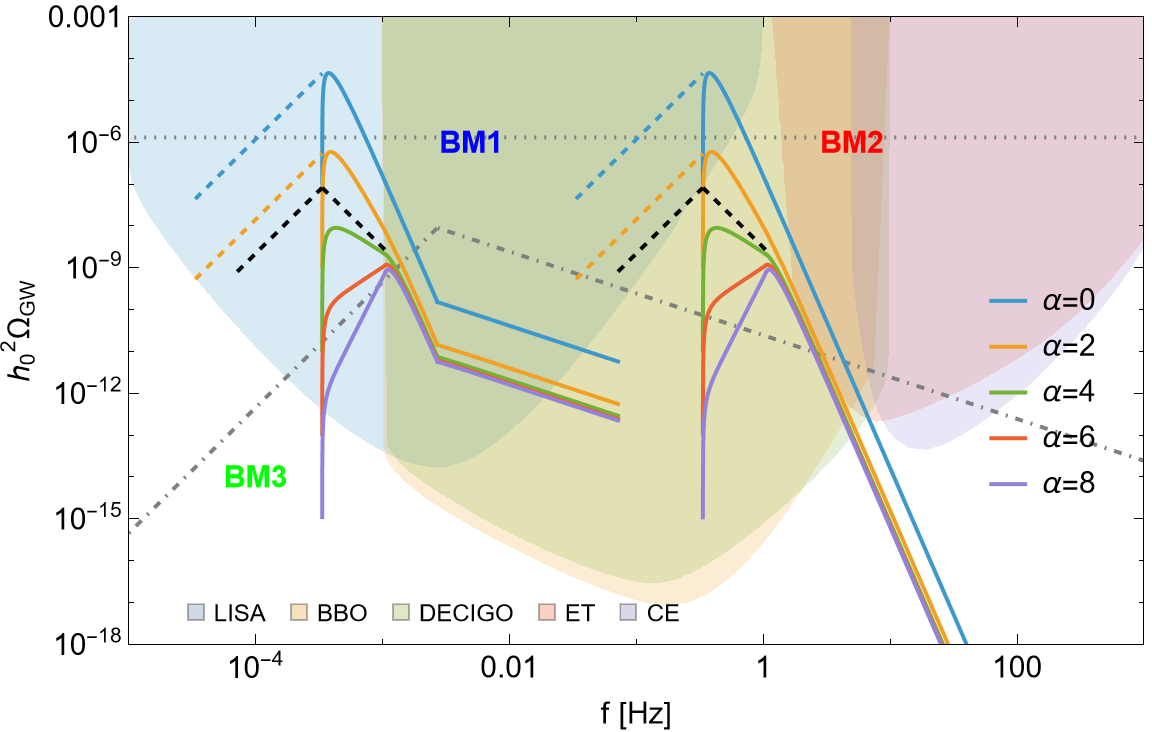}
    \caption{
    The GW spectrum with a DWD phase for two benchmarks, (BM1) and (BM2).
    We set $\nu = 2$ and $ r = 0.1$ for both. The dashed lines correspond to causality tails.
    The horizontal gray dotted line represents the $ \Delta N_{\rm eff}$ bound from CMB/BBN.
    The colored regions represent observational prospects from future GW experiments: LISA \cite{Audley:2017drz,Baker:2019nia}, DECIGO \cite{Seto:2001qf,Kawamura:2006up,Yagi:2011wg,Isoyama:2018rjb}, BBO \cite{Crowder:2005nr,Corbin:2005ny,Harry:2006fi}, Cosmic Explorer (CE) \cite{LIGOScientific:2016wof,Reitze:2019iox}, Einstein Telescope (ET)\cite{Punturo:2010zz,Hild:2010id,Sathyaprakash:2012jk,ET:2019dnz}. 
    The dot-dashed gray line corresponding to the case without DWD (with the choice (BM3)) is also shown for comparison. 
    \label{fig:example}}
\end{figure}

In Fig.~\ref{fig:example}, we plotted the GW spectra for two benchmarks with DWD, and one benchmark without DWD for comparison.
We took $ \nu = 2 $ and $r = 0.1$, and various choices of $ \alpha \in \{ 0 \,, 2 \,, 4\,, 6\,, 8 \}$.
The results for the general $ \nu $ and the opposite case of $ (k_{\rm weq} < ) \, k_{D} < k_{r}$ are presented in App.~\ref{supp:GW spectrum_other regime}.

A central result of this work is that the final GW spectrum is approximately independent of the wall tension $\sigma$.
To understand the $\sigma$-dependence, note that a larger DW tension $ \sigma$ enhances the magnitude of the GW power emitted.
At the same time, this shifts the onset of the DWD phase to an earlier time, leading to more dilution, causing a naive suppression of the signal. 
Interestingly, once the superhorizon evolution is properly taken into account, these effects cancel out, resulting in a $ \sigma$-independent GW spectrum. 
This is one of the characteristic and robust features of this scenario.
Additional details, as well as a more conceptual understanding of the parametric dependence of $\mathcal{A}$ can be found in App.~\ref{supp:back of envelope cal}.

Let us provide an intuitive explanation for the overall $k$-dependence appearing in Eq.~\eqref{eq:spectrum_2}. 
For $ k > k_{D} $, GWs evolve entirely on subhorizon from their emission until today, thereby preserving the spectral shape identical to RD only case, i.e.~$ \Omega_{\rm GW} \propto k^{1-\nu} $.
For $ k_{r} < k < k_{D} $, the spectrum is mainly dominated by the signal emitted at the latest time, $ t = r t_{\rm weq} (> t_{D}^{*} )$.
In this case, the transfer function is 
$\mathcal{T} \propto \frac{a_{e}}{a_{0}} \cdot \frac{a_3}{a_2} $
and $  a_3 / a_2 = ( k_{D}/k )^{3}  $ (see Eq.~\eqref{eq:transfer_2}),
so there is an additional $ k $ dependence coming from $\vert \mathcal{T} \vert^{2} \propto k^{-6} $. As a result, we obtain $ \Omega_{\rm GW} \propto k^{1-\nu} \cdot k^{-6} = k^{-5-\nu} $ as shown in Eq.~\eqref{eq:spectrum_2} for $ \nu = 2$.
Spectra for $ k < k_{r} $ show a strong dependence on the choice of $ \alpha $, and these IR modes are also bounded by the causality constraint.
The shape of the causality tail depends on the background evolution \cite{Hook:2020phx}. 
In deep IR regime, $ k < k_{\rm weq} $, modes stay extensively on superhorizon and enter the horizon later during RD after DW decay.
Therefore, these modes follow $ k^{3} $ profile \cite{Caprini:2009fx,Cai:2019cdl,Hook:2020phx}.
On the other hand, modes with $ k_{\rm weq} < k < k_{r} $ undergo a non-trivial evolution history, and the spectral shape also gets modified to $k^{-3}$, as explicitly shown in Sec.~\ref{sec:causality_tail}.
Fig.~\ref{fig:example} shows that, for $ \alpha  > 7/2 $, the low-frequency part of the GW spectrum
is dominated by the irreducible causality-tail envelope associated with the GW
emitted at $t=r t_{\rm weq}$.
Here the causality tail should be understood not as an additional independent contribution to be added on top of the plotted signal, but as an irreducible
infrared envelope, or lower bound, implied by causality.

Using the determined spectrum, we are able to constrain the possible parameter spaces by recasting observation data from CMB \cite{BICEP:2021xfz}, and PTA \cite{NANOGrav:2023gor,EPTA:2023fyk}.
For CMB, we use the tensor-to-scalar ratio $ r<0.036 $ \cite{BICEP:2021xfz} to extract the constraint at the CMB pivot scale $ k = 0.05 \, \text{Mpc}^{-1} $ (corresponding to $ f_{\rm CMB} \simeq 7.7 \times 10^{-17} \, \text{Hz} $) interpreted as $h^{2} \Omega_{\rm GW} (f_{\rm CMB} ) \lesssim 1.3\times 10^{-16} $  \cite{Maggiore:2018sht}.
However, parameter space constrained by CMB is already ruled out due to large inhomogeneities induced on CMB and LSS. See App.~\ref{supp:Fluctuation from DW Networks} for details.
For PTA, as a proxy for the frequency range and amplitudes PTA experiment covers, we impose  $ h^{2} \Omega_{\rm GW} (f_{\rm PTA}) \lesssim 2.5 \times 10^{-15} $ at $ f_{\rm PTA} = 1 \, \text{yr}^{-1} \simeq 3.2 \times 10^{-8} \, \text{Hz}$.\footnote{In fact, because PTA announced the discovery of SGWB, rather than simply treating this as a constraint, there are opportunities to explain the observed spectrum with the causality tail with $\sim k^{3}$.}
In Fig.~\ref{fig:inhom}, by taking $ \nu = 2 $ and $ \alpha = 7/2$ and $ r = 0.1 $, we present the parameters constrained by CMB (blue), PTA (orange), and large-scale inhomogeneities (green).
For $ \alpha =0 $ with other parameters taken the same as before, constraints in principle become slightly tighter and are shown as extended solid lines, but overall the difference is not significant.
We also show the GW bound associated with $\Delta N_{\rm eff} \lesssim 0.2$ \cite{Planck:2018vyg}, for
which the relevant constraints come from both BBN and CMB. This corresponds to gray dashed lines in Fig.~\ref{fig:example}.

\section{Conclusion and discussions}
\label{section:conclusion_and_discussions}

In this work, we consider the existence of a temporary phase of domain wall dominance (DWD) in the early universe, and study the characteristics of stochastic gravitational waves (GW) associated with such a DWD phase.
While long-lived domain walls are widely overlooked since they are typically considered to be problematic, i.e.,~domain wall problem, we show that it may be not only compatible with existing observations, but also reveal several remarkable features, including the conservation of tensor perturbations due to the superhorizon evolution during DWD phase, which are imprinted on the GW spectrum.
Incorporating unique features of DWD phase, we have derived detailed predictions for the GW spectrum, including its amplitude and spectral shape.
Among the most robust outcomes is the approximate independence of the final GW signal from the wall tension $\sigma$, which follows from the interplay between enhanced GW emission, the earlier onset of DWD, and the modified superhorizon evolution.

Combined with the multitude of theories predicting domain walls in the early universe, and the fact that DWD is very natural to occur, our results highlight the importance of domain wall dominance as a potential source of observable GW signals distinct from other astrophysical and cosmological processes.
We provide the first testable predictions for GW with an early DWD phase on upcoming GW detectors, which may probe this exciting possible novel phase of our universe.

\acknowledgments

The authors thank Chiara Caprini, Valerie Domcke, Gabriele Franciolini, Heejoo Kim, Hitoshi Murayama, Fabrizio Rompineve, and Liantao Wang  for inspiring discussions.
We also thank the anonymous referee for helpful comments and suggestions.
The work of SH and SML is supported by the National Research Foundation of Korea (NRF) Grant RS-2023-00211732, by the Samsung Science and Technology Foundation under Project Number SSTF-BA2302-05, and by the POSCO Science Fellowship of POSCO TJ Park Foundation. 
The work of SH is also supported by the National Research Foundation of Korea (NRF) Grant RS-2024-00405629.
The work of SML is also supported by the National Research Foundation of Korea (NRF) Grant 2012K1A3A2A0105178151. QL is supported by World Premier International Research Center
Initiative (WPI Initiative), MEXT, Japan.

\appendix

\section{VOS Model}
\label{supp:VOS Model}

Velocity-dependent One-Scale (VOS) model is first introduced for cosmic strings and captures the scaling behavior in RD and MD \cite{Martins:1995tg,Martins:1996jp,Martins:2000cs,Sousa:2013aaa}.
For the DW network, it was adopted in \cite{Kawano:1989mw,Avelino:2005kn,Leite:2011sc}.
We will use VOS model for DW with some extrapolations, to infer the behavior of the DW-network beyond RD/MD, even when DW energy density dominates the universe.
Two governing equations are
\begin{align}
    \frac{d v_{w}}{dt} & = (1-v_{w}^{2}) \left( \frac{k_{w}}{L} - 3 H v_{w} \right) \, . \label{eq:vos_veolocity}
\end{align}
and
\begin{align}
    \frac{d\rho_{w}}{dt} + H (1 + 3 v_{w}^{2}) \rho_{w} = - \frac{c_{w} v_{w}}{L} \rho_{w} - 
    p_{\rm GW}\label{eq:energy_veolocity}
\end{align}
where $k_w$ parametrizes the acceleration induced by wall curvature, $c_w$
parametrizes energy loss from intercommunication among DWs, $v_{w} $ is root-mean-squared (rms) velocity of the wall, and $ L $ is a typical curvature radius scale $ L = \sigma / \rho_{w} $, with $ \sigma $ and $ \rho_{w} $ being the surface tension and the energy density of the domain wall, respectively.
Terms proportional to $ H $ represent the dilution (for both velocity and energy) due to the Hubble expansion.
We also added $ p_{\rm GW} $-term on the right-hand side in Eq.~\eqref{eq:energy_veolocity} to take into account the energy (power) loss through gravitational wave production. This is mostly negligible except at onset of the domain wall dominance.
$ c_{w} $ and $ k_{w} $ should be calibrated with the numerical simulations.
Results from simulations performed assuming MD/RD universes tell us that $ (c_{w}, k_{w}) \simeq (0.5, 1) $ \cite{Avelino:2005kn,Leite:2011sc}.
We note that  this may not be true for DWD era, but we do not think this changes our results significantly because we are mainly concerned with the GW production during the scaling regime as explained in Sec.~\ref{sec:GW spectrum}, i.e.~GW production in DWD-phase is suppressed. Also, in general, $ c_{w} $ is believed to be a constant, while $ k_{w} $ may have a mild $ v_{w} $ dependence \cite{Avelino:2005kn}.

Under the assumption that the typical distance among DWs is also given by the same scale $ L $ (hence justifying the name `one-scale' model), i.e. $ \rho_{w} = \sigma /  L $, we get an equation for $L$:
\begin{align}
    \frac{dL}{dt} & = ( 1 + 3 v_{w}^{2} ) H L + c_{w} v_{w} + \frac{L^{2}}{\sigma} \frac{d\rho_{w}}{dt} \,. 
\end{align}
It may be worth mentioning that we have neglected effects of the frictional force caused by the particle scattering. This is model-dependent and it is likely that addition of friction will not change the overall conclusion significantly.
For a decelerating power-law expansion of the universe including RD and MD universe, $
    a(t) \propto t^{\lambda} $ with $ 0 < \lambda < 1 $, 
there exists an attractor solution $ L = \xi t $ with $ \xi \sim \mathcal{O}(1) $ and constant $ v_{w} $.
We note that, at earlier stages of scaling regime, we can neglect GW production captured by $p_{\rm GW}$-term.
Therefore, with $L \sim t \sim H^{-1}$, DW world-volume within a Hubble time per Hubble volume $n_{\rm DW} \sim L^2 t / H^{-3} \sim\mathcal{O} (1)$ during RD and MD, i.e. on average we have $\mathcal{O}(1) $ DW in a Hubble volume.


\section{Propagation of Tensor Perturbation}
\label{supp:tensor_pert}

For a tensor perturbation of the metric $ g_{\mu\nu} = \eta_{\mu\nu} + h_{\mu\nu} $ with $ \vert h_{\mu\nu} \vert \ll 1 $, the evolution equation is given by 
\begin{align}
    h ^{\prime\prime} + 2 \mathcal{H} h^{\prime} - \nabla^{2} h = a^{2} \frac{ 32 \pi G  \rho}{3} \Pi^{\rm TT}
    \label{eq:tensor_pert}
\end{align}
where we suppressed the tensor index denoting the polarization, and $ \prime $ denotes the derivative with respect to the conformal time $ \eta = \int dt / a(t) $. Also, $\mathcal{H} \equiv a^{\prime} / a $, and $\Pi_{\lambda}^{\rm TT}$ is the transverse, traceless part of the anisotropic stress, which is the only part in a source term relevant for the generation of GW.

For its propagation, we can set $ \Pi^{\rm TT} = 0 $.  Defining $ f \equiv a(\eta) h $, the evolution equation simplifies to
\begin{align}
    f^{\prime\prime} + \left( k^{2} - \frac{a^{\prime\prime}}{a} \right) f = 0 \, .
\end{align}
Therefore, for subhorizon mode ($ k \gg \mathcal{H} $) 
\begin{align}
    h (a(\eta), k) = \frac{1}{a(\eta)} \left( c_{1} e^{i k \eta} + c_{2} e^{-i k \eta} \right) \propto a^{-1}
\end{align}
while for superhorizon mode ($ k \ll \mathcal{H} $),
    \begin{align}
        h (a(\eta), k) = c_{1} + c_{2} \int \frac{1}{a^{2}(\eta)} \propto a^{0}
    \end{align}
where we have neglected the decaying part to obtain an overall scaling. Hence, we see that the amplitude is frozen on superhorizon, while it decays as $\propto a^{-1}$ on subhorizon. In the main text and in Sec.~\ref{sec:Characteristic_Scales_Transfer_Functions}, these facts are used to derive transfer functions $ \mathcal{T} $ depending on comoving momentum and the GW production time.

\section{Derivation of Characteristic Scales}
\label{supp:Derivation_of_Characteristic_Scales}

In this appendix, we provide details of the derivation of the characteristic scales
used in the main text, $k_D$ and $k_{\rm weq}$, given in Eq.~\eqref{eq:kWD}.

\paragraph{Derivation of $ k_{D}$}

By definition, $k_D \equiv a_D H_D$. Assuming instantaneous DW decay into radiation at $T=T_D$, we have
\begin{align}
    \rho_{r}(T_{D}) = 3 M_{P}^{2} H_{D}^{2} = \frac{\pi^{2}}{30} g_{*}(T_{D}) T_{D}^{4} \,, && H_{D} = \sqrt{ \frac{\pi^{2} g_{*}(T_{D})}{90} } \frac{T_{D}^{2}}{M_{P}} \,.
\end{align}
On the other hand, entropy conservation implies
\begin{align}
    \frac{a_{D}}{a_{0}} = \frac{T_{0}}{T_{D}} \left( \frac{g_{*s}(T_{0})}{g_{*s}(T_{D})} \right)^{1/3} \,.
\end{align}
Setting $a_0=1$, we obtain
\begin{equation}
    \begin{aligned}
    k_{D} & = T_{0} \left( \frac{g_{*s}(T_{0}) }{g_{*s}(T_{D}) } \right)^{1/3}  \sqrt{ \frac{\pi^{2} g_{*}(T_{D})}{90} } \frac{T_{D}}{M_{P}} \\
    & \simeq 1.7 \times 10^{-7} \left( \frac{g_{*}(T_{D})}{106.75} \right)^{1/6} \left( \frac{T_{D}}{\rm GeV} \right) \, {\rm Hz}
    \end{aligned}
\end{equation}
where in the second line we used $g_{*s}(T_D) \simeq g_{*}(T_D)$, appropriate at high temperature.

\paragraph{Derivation of $k_{\rm weq}$}

The time $t_{\rm weq}$ is defined by $\rho_w(t_{\rm weq})=\rho_r(t_{\rm weq})$.
Using the RD expressions, we have
\begin{align}
    \rho_{w}(t_{\rm weq})  = \frac{\sigma}{t_{\rm weq}} \, && \rho_{r}(t_{\rm weq}) =  \frac{3M_{P}^{2}}{4t_{\rm weq}^{2}} \,, && t_{\rm weq} = \frac{3M_{P}^{2}}{4\sigma} \,.
\end{align}
At equality, 
\begin{align}
    H_{\rm weq} = \frac{1}{2 t_{\rm weq}} = \frac{2\sigma}{3M_{P}^{2}} \,.
\end{align}
On the other hand, during DWD, $ \rho_{w} \propto a^{-1} $, so that
\begin{align}
   \frac{a_{\rm weq}}{a_{D}} = \frac{ \rho_{w}(t_{D}) }{ \rho_{w}(t_{\rm weq}) } = \frac{\pi^{2}}{30} g_{*}(T_{D}) T_{D}^{4} \frac{t_{\rm weq}}{\sigma} = \frac{\pi^{2}}{40} g_{*}(T_{D}) \frac{M_{P}^{2}T_{D}^{4}}{\sigma^{2}}  \,.
\end{align}
Therefore,
\begin{equation}
    \begin{aligned}
    k_{\rm weq} & \equiv a_{\rm weq} H_{\rm weq} = a_{D} \frac{a_{\rm weq}}{a_{D}} H_{\rm weq} \\
    & = \frac{\pi^{2}}{60} T_{0} g_{*,D} \left( \frac{g_{*s}(T_{0})}{g_{*s}(T_{D})} \right)^{1/3} \frac{T_{D}^{3}}{\sigma} \\
    & \simeq 2.1 \times 10^{12} \left( \frac{g_{*}(T_{D})}{106.75} \right)^{2/3} \left( \frac{T_{D}^{3}}{\sigma} \right) \, {\rm Hz} \,.
\end{aligned}
\end{equation}

\section{Fluctuation from DW Networks}
\label{supp:Fluctuation from DW Networks}

The existence of DWs in the early universe can also generate scalar perturbations, potentially leading to large inhomogeneities, which may be constrained by the CMB or large-scale structure (LSS). In this section, we briefly discuss these aspects.\footnote{Fluctuations from DW networks could also result in primordial black hole production \cite{Ferrer:2018uiu,Gelmini:2022nim,Gelmini:2023ngs,Gouttenoire:2023ftk,Gouttenoire:2023nzr,Gouttenoire:2023gbn,Dunsky:2024zdo,Lu:2024ngi} while the evolution of the scalar fluctuation associated with domain wall dominance has not been studied.
It is of interest to understand how these considerations are modified in the DWD scenario.}

Let us assume that the number density fluctuation of domain wall follows Poisson distribution at the time of its decay $ t_{D} $. Then probability of finding $ n $ walls in a volume $ V = R^{3} $ with some length scale $ R $ is given by
\begin{align}
   \mathbb{P} (n, R) = \frac{(\bar{n} R^{3})^{n}}{n!} \exp \left( - \bar{n} R^{3} \right)
\end{align}
where $ \bar{n} $ is the average comoving number density. At the time of its decay, it would be
\begin{align}
    \bar{n} \sim (a_{\rm weq} H_{\rm weq})^{3} \left( \frac{a_{\rm weq}}{a_{D}} \right) \simeq (a_{\rm weq} H_{\rm weq})^{3} \left( \frac{k_{\rm weq}}{k_{D}} \right)^{2}
\end{align}
where we assumed the value of the scaling regime (roughly one DW per one Hubble volume) at the beginning of DWD era, and then took into account the dilution during DWD $\propto a^{-1}$.

The variance in the number density is given by
    \begin{align}
        \sigma_{n}(R)^{2} = \langle (n - \bar{n} )^{2} \rangle_{R}
        =  \frac{\bar{n}}{R^{3}}
\end{align}
Therefore, as we consider larger volume (i.e.~larger $R$), the distribution is more coarse-grained, hence variance decreases.

Then, the energy contrast becomes
\begin{align}
    \delta = \frac{\Delta \rho_{\rm tot}}{ \bar{\rho}_{\rm tot}} \simeq  \frac{\Delta \rho_{w}}{ \bar{\rho}_{w}}
\end{align}
where $ \rho_{\rm tot} = \rho_{r} + \rho_{w} $ and we assumed DW distribution is the only source of the fluctuation and total energy density is also dominated by $\rho_w$ at that time. Then, corresponding variance is given by
\begin{equation}
     \begin{aligned}
    \sigma_{\delta}(R)  & = \sqrt{ \langle \delta^{2} \rangle_{R} } = \frac{ \sqrt{  \langle  (\Delta \rho_{w})^{2}  \rangle_{R} } }{ \bar{\rho}_{w} } \simeq \frac{1}{\sqrt{\bar{n} R^{3}} }   \simeq  \left(   a_{\rm weq} H_{\rm weq}  R\right)^{-3/2} \left( \frac{k_{D}}{k_{\rm weq}} \right) 
    \end{aligned} \label{eq:variance}
\end{equation}
where we used $ \bar{\rho}_{w} = M_{w} \bar{n} $ and $ \sqrt{ \langle  (\Delta \rho_{w})^{2}  \rangle } = M_{w} \sqrt{ \langle  (\Delta n )^{2}  \rangle } = M_{w} \sqrt{\bar{n} / R^{3}} $. (See \cite{Lu:2024ngi} for the application of the similar consideration for PBH production in RD universe.)

In momentum space, let us take the power spectrum of $ \delta $ to be
\begin{align}
    \mathcal{P}_{\delta}(k)  =
        A \left( \frac{k}{k_{\rm UV}} \right)^{m}  \qquad (k \leq k_{\rm UV}) 
\end{align}
with some coefficient $A$  and power $m$ which will be determined soon and $ k_{\rm UV} \simeq  k_{D}  $ corresponding to the Hubble scale at the time of consideration \cite{Zeng:2023jut}, and we will neglect contributions from modes with smaller than $ k_{D} $. Introducing Gaussian window function,
\begin{equation}
    \begin{aligned}
    \sigma_{\delta}^{2} \vert_{t=t_{D}} = \int_{0}^{k_{\rm UV}} \frac{dk}{k} ~ P_{\delta}(k) \exp \left( - k^{2} R^{2} \right) 
    \simeq \frac{1}{2} \Gamma\left( \frac{n}{2} \right) A ( a_{D} H_{D} R )^{-m}  \,  
    \end{aligned}
\end{equation}
where $ \Gamma(x) $ is Gamma function, and we can match with Eq.~\eqref{eq:variance} to determine $m = 3$ as well as $ A $.
\begin{align}
    \mathcal{P}_{\mathcal{\delta}}(k)  = \begin{dcases}
        \frac{4}{\sqrt{\pi}}  \left( \frac{k_{D}}{k_{\rm weq}} \right)^{2}  \left( \frac{k}{k_{\rm weq}} \right)^{3}  & (k \leq k_{D}) \\
        0  & (k > k_{D})
    \end{dcases} \label{eq:dw}
\end{align}

From CMB and LSS observations, we can probe the range (see, for instance, \cite{Bringmann:2011ut})
\begin{align}
    k_{\rm CMB} \in ( 10^{-3}, \, 10^{-1})~\text{Mpc}^{-1}  \,,  && k_{\rm LSS} \in ( 10^{-1}, \, 3)~\text{Mpc}^{-1}    .
\end{align}
We have $ P_{\mathcal{R}} \simeq 2 \times 10^{-9} $ at these scales, with fluctuations at the level of
\begin{align}
    \frac{ \Delta \mathcal{P}_{\mathcal{R}} }{ \mathcal{P}_{\mathcal{R}} }  \lesssim \begin{dcases} 10^{-2}  & (\text{CMB}) \\
     10^{-1}  & (\text{LSS})
    \end{dcases}
\end{align}
where we take $ \Delta \mathcal{P}_{\mathcal{R}} \sim \mathcal{P}_{\mathcal{\delta}}(k) $ from domain walls, Eq.~\eqref{eq:dw}.

\section{General Formulas}
\label{sec:General_Formulas}

In this section, we provide more general formulas of the transfer functions $ \mathcal{T} $ and the GW spectrum beyond the ones presented in the main text.

\subsection{Transfer Functions}
\label{supp:Transfer Functions}

Below, we provide the complete list of transfer functions.

\begin{itemize}

\item $ t_{e} < t_{D}^{*} $

\begin{align}
    \mathcal{T}(a_{e},k) = 
    \begin{dcases}
       \frac{ a_{3} }{ a_{0} }   &     
 \left( k < k_{\rm weq} \right) \\
         \left( \frac{ a_{1} }{ a_{2} } \right) 
         \left( \frac{ a_{3} }{ a_{0} } \right)  &    \left( k_{\rm weq} \leq k < k_{D} \right) \\  
         \left( \frac{ a_{1} }{ a_{0} } \right)  &   ( k_{D} \leq k < a_{e} H_{e} ) \\
         \frac{ a_{e} }{ a_{0} }    
          &   \left(   a_{e} H_{e}  \leq k \right)
    \end{dcases}
\end{align}

\item $ t_{D}^{*} < t_{e} < t_{\rm weq}$ 

\begin{align}
    \mathcal{T}(a_{e},k) = 
    \begin{dcases}
       \frac{ a_{3} }{ a_{0} }   &
 \left( k < k_{\rm weq} \right) \\
         \left( \frac{ a_{1} }{ a_{2} } \right)
        \left( \frac{ a_{3} }{ a_{0} } \right)  &    \left( k_{\rm weq} \leq k <  a_{e} H_{e}  \right) \\
         \left( \frac{ a_{e} }{ a_{2} } \right)  
       \left( \frac{ a_{3} }{ a_{0} } \right)  &   \left( a_{e} H_{e} \leq k <   k_{D} \right) \\
           \frac{ a_{e} }{ a_{0} }  
          &   \left(  k_{D}  \leq k  \right)
    \end{dcases}
\end{align}
   
\item $  t_{\rm weq} < t_{e} < t_{D} $

\begin{align}
    \mathcal{T}(a_{e},k) = 
    \begin{dcases}
      \frac{ a_{3} }{ a_{0} }  &   
 \left( k < a H_{e} \right) \\
       \left( \frac{ a_{e} }{ a_{2} } \right) \left( \frac{ a_{3} }{ a_{0} } \right)  &   \left(  a_{e} H_{e}   \leq k <   k_{D} \right) \\
        \frac{ a_{e} }{ a_{0} } 
          &   \left(  k_{D}  \leq k \right)
    \end{dcases}
\end{align}

\end{itemize}

\subsection{GW Spectrum}
\label{supp:GW spectrum_other regime}

In the main text, we provide the form of GW spectrum for the case of $k_{\rm weq} < k_{r} < k_{\rm D}$ specifically choosing $ \nu = 2 $. Here, we provide the results for general $ \nu $ as well as the results for the other case of $ k_{\rm weq} < k_{D} < k_{r} $.

\begin{itemize}
    \item  $k_{\rm weq} < k_{r} < k_{D}$
\begin{equation}
    \begin{aligned}
    \footnotesize
     h^{2} \Omega_{\rm GW} (k)  
       \simeq  \begin{dcases}
        \frac{ 2r^{\alpha} (\nu-1) }{5-2\alpha-\nu}  \mathcal{A} \left[ \left( \frac{k}{ k_{\rm weq}} \right)^{-5-\nu}
       - \left( \frac{k }{ k_{\rm weq}} \right)^{-10+2\alpha} 
       \right] & ( k_{\rm weq} < k < k_{r}  ) \\
       \frac{2 r^{5} (\nu-1) }{5 - \nu} \mathcal{A} \left[   \frac{ \left( r^{\alpha + \frac{\nu}{2} } (5 - \nu) -  2 r^{\frac{5}{2}} \alpha \right) }{r^{\frac{5}{2}}(5 - 2 \alpha - \nu)}
       \left( 
       \frac{k}{k_{r}} \right)^{-5-\nu}
       - \left( \frac{k}{ k_{r}} \right)^{-10} 
       \right] & ( k_{r} < k < k_{D} ) \\
       \frac{2 (\nu-1) }{5 - \nu} \mathcal{A} \left[   \frac{ \left( r^{\alpha } (5 - \nu) -  2 r^{\frac{5}{2} - \frac{\nu}{2}} \alpha \right) }{  (5 - 2 \alpha - \nu)}
       \left(  \frac{k_{D}}{k_{\rm weq}}\right)^{- 5 - \nu}
       \left( 
       \frac{k}{k_{D}} \right)^{1-\nu} \right. \\ \left. \hspace{5cm} - \left( \frac{k_{D}}{k_{\rm weq}} \right)^{-10} \left( \frac{k}{ k_{D}} \right)^{-4} 
       \right] & (k>k_{D})
       \end{dcases}
\end{aligned}
\end{equation}

\item $ k_{\rm weq} < k_{D} < k_{r} $ 
\begin{equation}
\footnotesize
 \begin{aligned}
      h^{2} \Omega_{\rm GW} (k)  
       \simeq \begin{dcases}
           \frac{ 2r^{\alpha} (\nu-1) }{5-2\alpha-\nu}  \mathcal{A} \left[ \left( \frac{k}{ k_{\rm weq}} \right)^{-5-\nu}
       - \left( \frac{k }{ k_{\rm weq}} \right)^{-10+2\alpha} 
       \right] & (   k_{\rm weq} < k < k_{D} ) \\
        \frac{2 r^{\alpha} (\nu-1) }{5 - 2\alpha - \nu} \mathcal{A} \left[   
        \left( \frac{k_{\rm weq}}{k_{D}} \right)^{5 + \nu} \left( \frac{k}{k_{D}} \right)^{-\nu +1} - \left( \frac{k_{\rm weq}}{k_{D}} \right)^{10-2\alpha} \left( \frac{k}{k_{D}} \right)^{2\alpha - 4}
       \right] & (  k_{D} < k < k_{r}  ) \\
       \frac{2 r^{2} (\nu-1) }{5 - \nu} \mathcal{A} \left[   \left( \frac{k_{\rm weq}}{k_{D}} \right)^{6} \left( \frac{k}{k_{\rm p}}\right)^{4}  \right. \\ \left. \hspace{2.5cm}  - 2 \frac{2\alpha + r^{(2\alpha + \nu -5)/2} (\nu-5) }{2\alpha + \nu -5} \left( 
 \frac{k_{\rm weq}}{k_{D}}\right)^{6} \left( \frac{k}{k_{r}} \right)^{1-\nu}
       \right] & (  k > k_{r} )
       \end{dcases}
\end{aligned}   
\end{equation}

\end{itemize}

\begin{figure}
    \centering
    \includegraphics[width=0.75\linewidth]{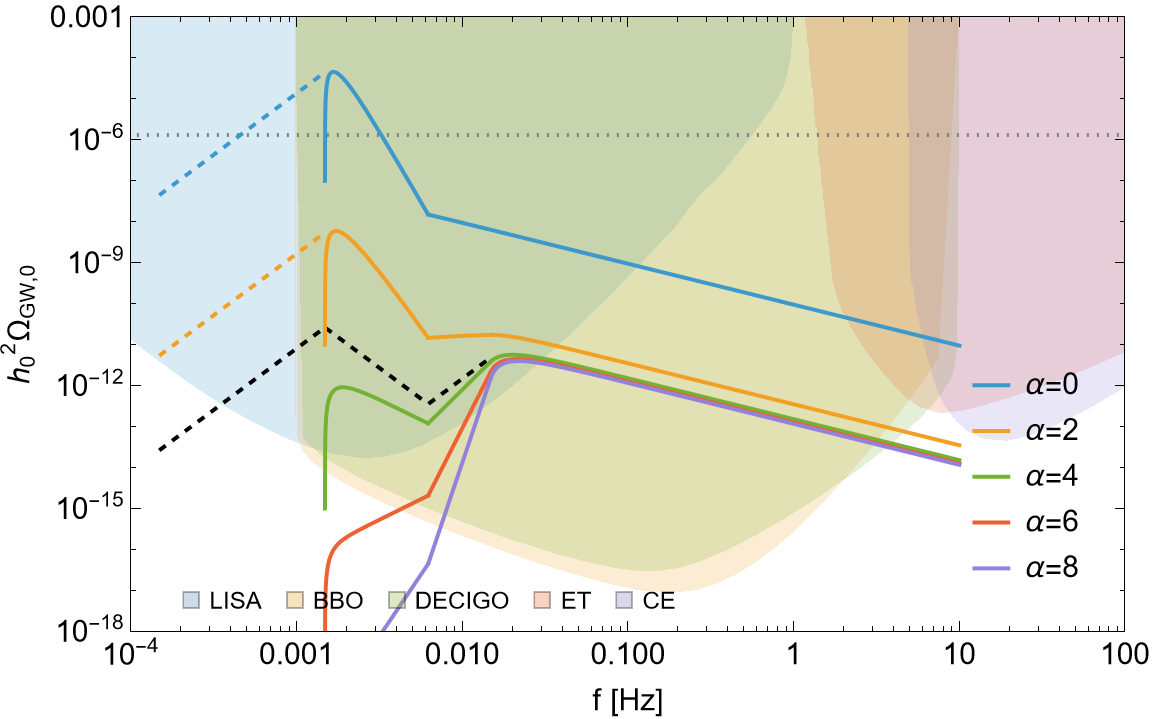}
    \caption{GW spectrum for the case 
    case of $  k_{\rm weq} < k_{D} < k_{r}   $, with parameters $\sigma = (1.4 \times 10^{10} \, \text{GeV})^{3} $, $T_{D} = 2.3\times 10^{5} \, \text{GeV}$, $r=10^{-2}$. Black dashed line corresponds to lower bound from the causality tail.}
    \label{fig:case2}
\end{figure}

In Fig.~\ref{fig:case2}, for the purpose of illustration, we present one case with choices $\sigma = (1.4 \times 10^{10} \, \text{GeV})^{3} $, $T_{D} = 2.3\times 10^{5} \, \text{GeV}$, and $r=10^{-2}$. 
For large $ \alpha $ values, there clearly exists significant change of the spectrum, including the spectrum of the causality tail depicted in gray, which may be compared to the one shown in the main text, Fig.~\ref{fig:example}.
For causality tail, modes with $ k < k_{\rm weq} $ and $ k_{D} < k < k_{r} $ undergo horizon crossing only once during RD (the first is after the domain wall decay, and the second is before the wall domination, respectively), therefore the causality tail follows $ k^{3} $ spectrum.
On the other hand, for the modes with $k$ in between, they do have $k^{-3}$ spectrum due to multiple horizon crossings and related partial subhorizon evolution.
As a results, for $ \alpha > 7/2 $, the GW spectrum is hardly sensitive to the values of $ \alpha $.

\section{Back-of-the-Envelope Calculation}
\label{supp:back of envelope cal}

In this section, we provide back-of-the-envelope calculation and explain the parametric dependence of the SGWB presented in the main text denoted by $ \mathcal{A} $.

To make the situation simpler, let us assume that the scaling solution holds all the way to the very vicinity of the domain wall dominance, equivalently setting $r = 1$. 
Then, we are mainly interested in the peak amplitude at $ k =  k_{\rm weq} $.
This mode enter the horizon again when the scale factor have the value
\begin{align}
a_{*} = a_{\rm eq} \left( \frac{k_{\rm eq}}{k_{\rm weq}} \right)  = a_{D} \left( \frac{a_{\rm weq}}{a_{D}} \right)^{-1/2} \,.
\end{align}

Using the quadrupole formula applied to the scaling regime, the GW energy density is estimated to be of order
\begin{align}
    \rho_{\rm GW,weq} \sim G \sigma^{2}
\end{align}
at the time of wall-radiation equality. Then, at the time of horizon re-entry, the energy density of GW is diluted as
\begin{align}
    \rho_{\text{GW},*} = \rho_{\rm GW,weq} \left( 
 \frac{a_{\rm weq}}{a_{ *}}\right)^{2} \,.
\end{align}
Here, we used that the amplitude of the fluctuation freezes during the superhorizon evolution, hence the GW energy density dilutes only as $\propto a^{-2}$. This can be seen from Eq.~\eqref{eq:Omega_GW_k} using $h \sim a^0$ on superhorizon.
On the other hand, After the horizon re-entry, $\rho_{\rm GW}$ redshifts as $\propto a^{-4}$ since $h \sim a^{-1}$ on subhorizon:
\begin{align}
    \rho_{\rm GW,eq} & = \rho_{\text{GW},*} \left( 
 \frac{a_{*}}{a_{\rm eq}}\right)^{4}  = \rho_{\rm GW,weq} \left( 
 \frac{a_{\rm weq}}{a_{*}}\right)^{2} \left( 
 \frac{a_{*}}{a_{\rm eq}}\right)^{4}   =  \rho_{\rm GW,weq} \left( \frac{t_{\rm weq}}{t_{\rm eq}} \right)^{2} \,. \label{eq:rho_GW,eq}
\end{align}
where we used
\begin{align}
    a_{D}  = a_{\rm eq} \left( \frac{t_{D}}{t_{\rm eq}} \right)^{1/2} \, , \;\;\;\;\;\;
    a_{\rm weq} = a_{ D} \left( \frac{t_{\rm weq}}{t_{D}} \right)^{2} = a_{\rm eq} \left( \frac{t_{D}}{t_{\rm eq}} \right)^{1/2} \left( \frac{t_{\rm weq}}{t_{D}} \right)^{2}
\end{align}
in the last equality.

It is noteworthy that the final expression Eq.~\eqref{eq:rho_GW,eq} is independent of the power $\lambda$ in the background evolution $ a \propto t^{\lambda} $ as long as $ \lambda > 1 $.
This guarantees that the amplitude of the peak is independent of when DWs decay into radiation.
This is consistent with the fact that superhorizon mode is not sensitive physics on subhorizon (e.g. decay of domain walls).
The information of $t_{D}$ is indirectly encoded in the observed peak frequency as it affects the way frequency redshifts.
The expression Eq.~\eqref{eq:rho_GW,eq} can be further simplified by using $ t_{\rm weq} \sim M_{P}^{2} / \sigma $, and we get
\begin{align}
    \rho_{\rm GW,eq}  \sim G \sigma^{2}  \left( \frac{ M_{P}^{2}}{ \sigma t_{\rm eq}} \right)^{2} = \frac{M_{P}^{2}}{t_{\rm eq}^{2}} \sim M_{P}^{2} H_{\rm eq}^{2}
\end{align}
so that
\begin{align}
    \rho_{\rm GW,0} \sim a_{\rm eq}^{4} M_{P}^{2} H_{\rm eq}^{2}.
\end{align}
We notice that this final answer is independent of the wall tension $\sigma$, as discussed in Sec.~\ref{subsec:GW spectrum}.
Finally, we get
\begin{align}
    \Omega_{\rm GW} & \sim \frac{\rho_{\rm GW,0}}{\rho_{c,0}} \sim \frac{a_{\rm eq}^{4}M_{P}^{2} H_{\rm eq}^{2}}{M_{P}^{2} H_{0}^{2}} \sim 
    \frac{ a_{\rm eq}^{4} H_{\rm eq}^{2} }{ H_{0}^{2} }
\end{align}
which explains the parametric dependence of the $ \mathcal{A} $ we introduced in Sec.~\ref{subsec:GW spectrum}.

\bibliographystyle{unsrt}
\bibliography{arxiv_v2}

\end{document}